\RequirePackage{fix-cm}
\documentclass[smallextended]{svjour3}       
\smartqed  
\usepackage{graphicx}
\usepackage{color}
\begin{document}

\title{Anomalous attenuation of piezoacoustic surface waves by liquid helium thin films
\thanks{This work was supported by the NSF (Grant no. DMR-1708331).}
}


\author{H. Byeon \and K. Nasyedkin \and J.R Lane \and L. Zhang \and N.R. Beysengulov \and \newline  R. Loloee \and J. Pollanen}


\institute{H. Byeon \at
              Department of Physics and Astronomy, Michigan State University, East Lansing MI 48824-2320, USA \\
              \email{byeonhee@msu.edu}           
}

\date{Received: date / Accepted: date}

\maketitle

\begin{abstract}
We report on the observation of an anomalously high attenuation of high frequency surface acoustic waves by thin films of liquid $^{4}$He. The piezoelectric acoustic waves propagate along the surface of a lithium niobate substrate, which is coated with varying amounts of liquid helium. When the thickness of the helium layer is much larger than the wavelength of the surface acoustic wave on the substrate its attenuation is dominated by the excitation of compressional waves into the liquid, in good agreement with theory and previous measurements. However, for sufficiently thin helium coverage, we find that the acoustic wave attenuation is significantly increased beyond that measured with the substrate submerged in bulk liquid. Possible mechanisms for this enhanced attenuation are discussed.

\end{abstract}

\section{Introduction}
\label{intro}
The propagation of high frequency surface acoustic waves (SAWs) on a crystalline piezoelectric substrate are exquisitely sensitive to their physical environment. For this reason they have proven to be a versatile tool for studying a wide variety of physical systems placed in close proximity to the surface wave. These range from molecular adsorbates\cite{Bal89,Roc09} to two-dimensional electron systems\cite{wix86,wil932,esk96,pol16,fri17,Lan18} to superconducting qubits\cite{Gus14,Man17}, just to name a few. SAWs have also found utility in studying quantum fluids where they have been used to explore the bulk acoustic properties of liquid $^{3}$He and $^{4}$He\cite{Arz67,Tok79,Che82,Aok03,Aok032} and to investigate superfluidity of extremely thin films (2-10 monolayers) of $^{4}$He\cite{Mor80,Che822,Kos90}. However, to our knowledge, no surface acoustic wave experiments have been performed on helium films having a thickness on the order of $\sim 10-100$ nm, where one would expect strong coupling of the SAW to excitations in the liquid helium surface. Moreover, since films of this thickness are routinely used as a substrate in experiments with electrons on helium\cite{And97,Mon04}, understanding the coupling of a piezoelectric surface wave to the helium film is a prerequisite for future SAW measurements of the non-degenerate two-dimensional electron system that can be created on the surface of the liquid. 

In light of this, we have performed pulsed measurements of the attenuation, $\alpha$, of high frequency SAWs on the surface of lithium niobate caused by thin films of liquid helium. We find that for superfluid films with a thickness of $\approx 60 - 70$ nm the SAW attenuation anamolously exceeds that of bulk helium at the same temperature. We speculate that this increased attenuation is due to the electrostrictive coupling between the piezoacoustic wave and excitations in the helium film surface.

\section{Experiment}
\label{expt}
\begin{figure*}
  \includegraphics{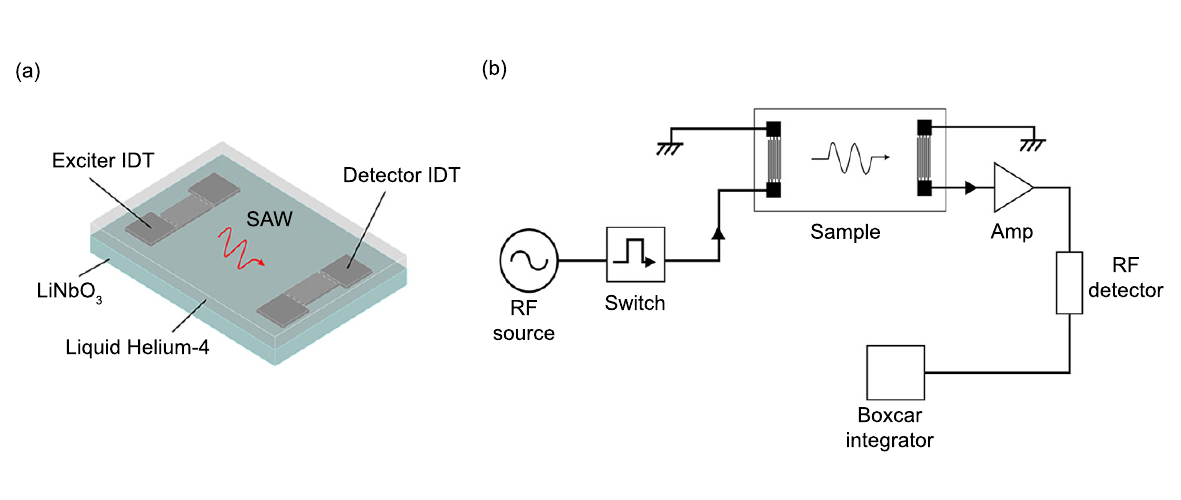}
\caption{Pulsed surface acoustic wave experimental setup. (a) Schematic of the lithium niobate chip used for exciting and detecting SAW via aluminium interdigitated transducers (IDT), which were directly fabricated onto the surface of the substrate. (b) Block diagram of the circuit used for measuring the attenuation of pulsed SAWs. As described in the text, by measuring the time-of-flight signal we are able to disentangle direct capacitive crosstalk between the two interdigitated transducers as well as multiply reflected SAW pulses.}
\label{fig1}
\end{figure*}

\begin{figure}
  \includegraphics{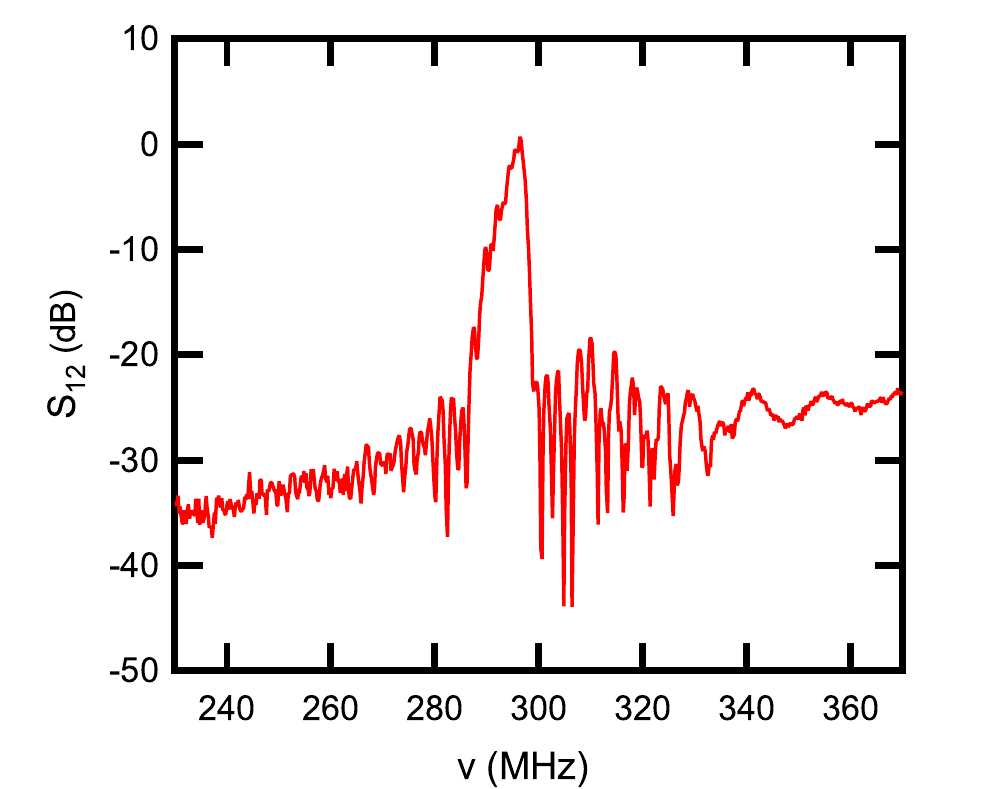}
\caption{Frequency dependence of the transmission coefficient ($S_{12}$) of the SAW delay line at $T = 1.55$ K in vacuum. The resonant peak at $\nu = 296$ MHz is associated with the generation of surface acoustic waves in the IDT delay line.}
\label{fig2}
\end{figure}

The SAW attenuation experiments reported here were performed using a YZ cut lithium niobate (LiNbO\textsubscript{3}) single crystalline wafer as a substrate for SAW propagation. The lithium niobate substrate was diced into a rectangle with a length of 20 mm and a width of 10  mm. To excite and detect SAWs, two identical interdigitated transducers (IDT) having 40 pairs of $3 \mu$m wide fingers were patterned on the surface a LiNbO$_{3}$ chip in the form of a delay line, having a length of 16.5 mm, along the crystallographic x-axis using conventional photo-lithography (see Fig.~\ref{fig1}a). The transducers were fabricated from aluminum and had a thickness of approximately 70 nm and a width of 4 mm defining the width of the SAW beam within the delay line. We note that in this geometry the IDTs are positioned 1.75 mm from the diced edge of the chip.
Resonant SAWs are launched by applying a high frequency signal between the transducer fingers of the exciter to create an elastic distortion of the piezoelectric substrate beneath the IDT. In this configuration, Rayleigh mode surface acoustic waves with components parallel (longitudinal) and perpendicular (transverse) to the wave propagation are launched along the LiNbO$_{3}$ surface toward the detector IDT. The fundamental resonant frequency of our Rayleigh wave SAW device is $\nu = v_{s}/\lambda = 291$ MHz, dictated by the IDT finger periodicity, $\lambda = 12~\mu$m, and the speed of sound in YZ-cut LiNbO$_{3}$, $v_{s} = 3488$ m/s. The frequency response of the SAW delay used in our attenuation measurements was characterized using an Agilent N5230A vector network analyzer. Fig.~\ref{fig2} shows the measured transmission coefficient, $S_{12}$, of the delay line as a function of frequency at $T = 1.55$ K in vacuum. 
The resonance in the transmitted power at $\nu \simeq$ 296 MHz is associated with the generation of SAW in the substrate. The measured resonance is a few percent larger than the expected value, likely due to an increase in the elastic moduli of LiNbO$_{3}$ at cryogenic temperature. 

Fig.~\ref{fig1}b shows a diagram of the circuit used to measure the attenuation of the piezoelectric waves. Pulsed SAWs are created by gating a continuous wave signal using a fast solid-state switch. For the measurements reported here the pulses had a width of $3~\mu$s and were repeated at a frequency of 1 kHz. The received SAW signal at the detector IDT was amplified by 30 dB and measured using a calibrated crystal diode detector (Krytar Model 203AK S/N 00277). A boxcar integrator was used for time-of-flight measurements. This mode of operation allows us to disentangle the piezoacoustic signal from direct capacitive crosstalk between the two IDTs as well as multiply reflected surface acoustic wave pulses within the delay line. 

To study the surface acoustic wave attenuation in liquid $^{4}$He, the SAW device was mounted inside of a hermetically sealed copper cell attached to a closed cycle 1K cryostat. Helium gas was supplied into the cell at T $\simeq$ 1.55 K through a capillary fill line. A resistive thermometer located outside of the cell and calibrated relative to the known superfluid transition temperature of $^{4}$He was used to measured the temperature of the liquid helium. The liquid helium volume admitted into the cell was determined by varying the pressure in a calibrated standard volume of 260 cc at room temperature. The thickness of the helium film can be estimated from\cite{Pob92},
\begin{equation}
d = \Big(\frac{\gamma}{\rho g H}\Big)^{1/4},
\end{equation}
where $\gamma$ is the van der Waals constant, $\rho$ is the mass density of $^{4}$He, $g$ is the acceleration due to gravity, and $H$ is the distance from the LiNbO$_{3}$ surface down to the liquid helium level in the reservoir volume in the cell. Three-dimensional modeling of the experimental cell open volume was used to calculate $H$ from the volume of helium admitted into the cell from the calibrated volume at room temperature.

\section{Results and Discussion}
\label{RandD}
Before presenting our results on thin liquid helium films, we first demonstrate that we are able to use our measurement setup to reproduce the known temperature dependence of Rayleigh wave attenuation by bulk liquid helium. Rayleigh mode surface acoustic waves are attenuated by contact with a bulk liquid due to the surface-normal component of the particle motion and that of the co-propagating piezoelectric field. These components generate longitudinal compressional waves upward into the liquid, dissipating most of the acoustic energy\cite{Arz67,Dran70}. This method of energy loss is closely related to the problem of the Kaptiza resistance based on the acoustic mismatch between liquid helium and a solid substrate\cite{Pob92,Pol09}, whereby longitudinal waves having a velocity $v$ are emitted into the liquid from the substrate surface at an angle $\phi = \arcsin(v/v_{s})$. In addition to this mechanism, the in-plane shear component of the SAW can also radiate energy into a fluid over the viscous penetration depth,
\begin{equation}
L = \sqrt{\frac{\eta}{\pi \nu \rho}}
\end{equation}
where $\eta$ and $\rho$ are the viscosity and mass density of the fluid. However, SAW energy lost to shear is two to three orders of magnitude smaller than longitudinal absorption in our temperature range and is negligible even for our measurements in thin helium films. 

Our results for the temperature dependence of the SAW attenuation produced by bulk liquid helium are shown in Fig. \ref{fig4} as the solid red curve. For comparison we also show the prediction for the attenuation (dashed red curve) based on the loss of SAW energy in the form of longitudinal compressional waves\cite{Dran70}. As expected, our measurements are in good correspondence with theoretical prediction. Moreover, our results for bulk helium are also in good agreement with previous measurements of the attenuation of piezoelectric Rayleigh waves by liquid helium\cite{Arz67,Tok79,Che82,Aok03}. 
\begin{figure}
  \includegraphics{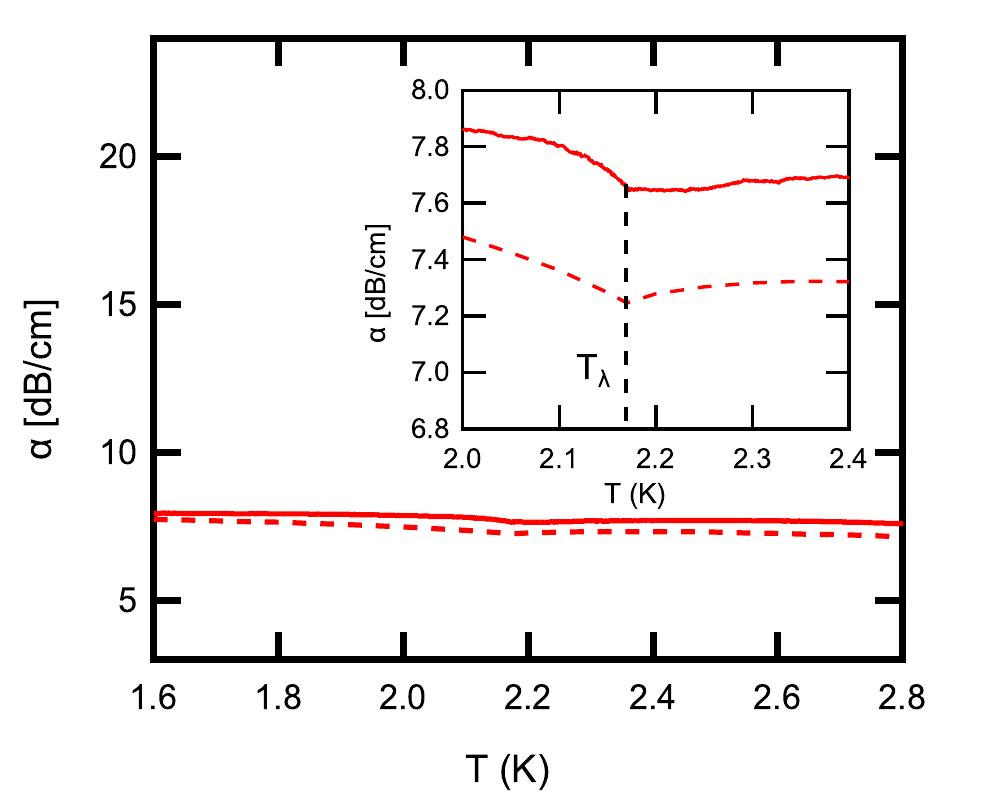}
\caption{Attenuation, $\alpha$, of Rayleigh waves on LiNbO$_{3}$ in contact with bulk liquid helium. Our measurements (solid red curves) at $\nu=296$ MHz are in good agreement with the theory (dashed red curve) of Dransfeld and Salzmann\cite{Dran70} for energy loss due to radiation of longitudinal compressional waves into the liquid. (Inset) The local minimum of the SAW attenuation at $T_{\lambda}= 2.17$ K is associated with the transition from normal to superfluid $^{4}$He.}
\label{fig3} 
\end{figure}

Unexpectedly, however, we find that when the thickness of the liquid helium layer is sufficiently reduced an anomalously large attenuation can be induced. In Fig.~\ref{fig4} we show measurements of the SAW attenuation (solid blue data) at $T = 1.55$ K while increasing the volume of helium in the cell by small increments. 
\begin{figure}
  \includegraphics{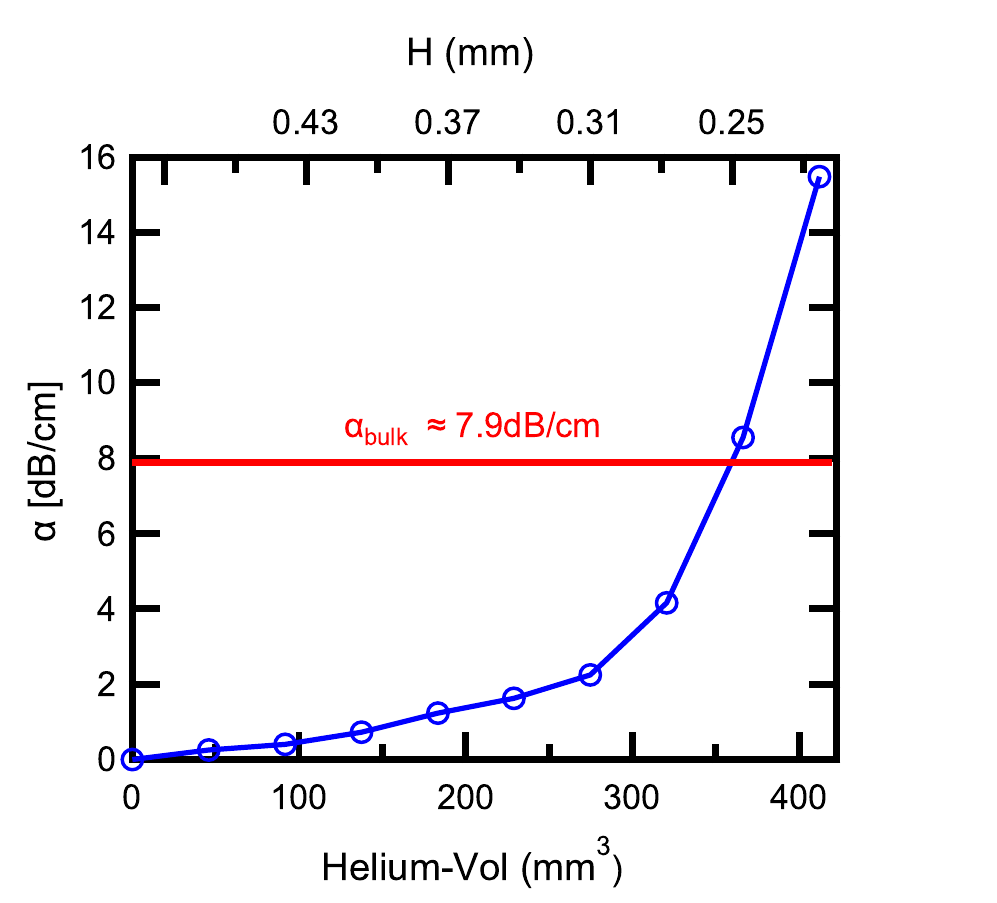}
\caption{Attenuation, $\alpha$, of Rayleigh waves on LiNbO$_{3}$ in contact with thin films of liquid helium at $T = 1.55$ K and $\nu = 296$ MHz. The measurements are made by incrementally admitting small amounts of helium gas into the cell from a standard volume at room temperature (lower horizontal axis). The vertical dashed line indicates the estimated thickness of the helium film on the LiNbO$_{3}$ surface based on Eq.~ 1 and the height $H$ from the substrate to the bulk helium reservoir level in the cell (top horizontal axis). For comparison, the horizontal solid red line indicates the SAW attenuation due to bulk liquid helium at the same temperature. The calculated thickness of the helium film for the final two data points above the bulk value are 73 nm and 76 nm. Finally, we note that the IDT fingers are expected to fill completely with helium via capillary action for very small amounts of helium introduced into the cell, \emph{i.e.} already at H = 0.46 mm}
\label{fig4}
\end{figure}
With an increasing amount of helium we observe a rapid increase in the SAW attenuation once the thickness of the superfluid film reaches $\approx$ 60 - 70 nm ($H \approx 0.28$ mm) and continues to increase and quickly exceeds the attenuation produced by bulk helium.

For reference we plot the SAW attenuation measured in bulk helium at the same temperature as the red horizontal line in Fig.~\ref{fig4}. We emphasize that this anomalously large attenuation is reproducible on multiple fillings of the experimental cell and additionally is insensitive to the tilt of the cryostat with respect to vertical\footnote{With further increasing film thickness the SAW attenuation will eventually recover the value measured for bulk helium. However in this crossover regime ($d \sim 1-10 \mu$m) the film thickness on the substrate is exceedingly sensitive to fluctuations in $H$ and also to the tilt of the cryostat leading to non-reproducibility. Future experiments, using microchannel geometries could allow for controlled measurement in this regime of film thickness.}. Moreover, we find that this phenomenon is not restricted to superfluid helium but rather persists into the normal state.

To our knowledge, no prior measurements of the attenuation of high frequency SAW have been made in this regime of helium thickness nor is there a theory describing the coupling of SAWs to such films. We speculate that the occurrence of the attenuation anomaly reported here for thin helium films can qualitatively be understood in terms of a coupling between the piezoelectric surface wave on the LiNbO$_{3}$ substrate and excitations in the helium surface. While the physical displacement of atoms associated with the SAW is at the angstrom scale, the SAW piezoelectric field has a spatial extent on the order of the SAW wavelength, which in our case is $\simeq 12~\mu$m. In fact, it is well-known that an electric field gradient can exert a force to move liquid helium via electrostriction, an effect which has even been utilized in developing a number of recent superfluid optomechanical systems\cite{Har16,Lor17,Sou17,Chi17}. It is possible that an electrostrictive coupling between the LiNbO$_{3}$ surface wave and the liquid helium surface conspire to produce a new mode of energy loss not present in bulk helium. Finally, we note that our observation of enhanced attenuation in thin \emph{normal} helium films is able to rule out the coupling of SAWs to third sound as the sole mechanism for increased SAW energy loss.

\section{Conclusion}
\label{conclude}
In conclusion, we have performed high frequency SAW attenuation measurements in thin films of superfluid and normal $^{4}$He where we find an anomalously large loss of energy from the piezoelectric surface wave into the liquid. We suspect that this increased attenuation is associated with electromechanical excitation of the helium film surface via a coupling of the SAW electric field to the dielectric constant of liquid helium.

\begin{acknowledgements}
We are grateful to M.I. Dykman and W.P. Halperin for helpful discussions. This work was supported by the NSF (Grant no. DMR-1708331).
\end{acknowledgements}

\bibliographystyle{spphys}

\begin{thebibliography}{10}
\providecommand{\url}[1]{{#1}}
\providecommand{\urlprefix}{URL }
\expandafter\ifx\csname urlstyle\endcsname\relax
  \providecommand{\doi}[1]{DOI \discretionary{}{}{}#1}\else
  \providecommand{\doi}{DOI \discretionary{}{}{}\begingroup
  \urlstyle{rm}\Url}\fi

\bibitem{Bal89}
D.~Ballantine, H.~Wohltjen, Analytical Chemistry \textbf{61}(11), 704A (1989)

\bibitem{Roc09}
M.~Rocha-Gaso, C.~March-Iborra, A.~Montoya-Baides, A.~Arnau-Vives, Sensors
  \textbf{9}, 5740 (2009)

\bibitem{wix86}
A.~Wixforth, J.P. Kotthaus, G.~Weimann, Phys. Rev. Lett. \textbf{56}, 2104
  (1986)

\bibitem{wil932}
R.L. Willett, R.R. Ruel, K.W. West, L.N. Pfeiffer, Phys. Rev. Lett.
  \textbf{71}, 3846 (1993)

\bibitem{esk96}
M.~Eskildsen, H.~Smith, J. Phys. Condens. Matt. \textbf{8}, 6597 (1996)

\bibitem{pol16}
J.~Pollanen, J.~Eisenstein, L.~Pfeiffer, K.~West, Phys. Rev. B \textbf{94}(24),
  245440 (2016)

\bibitem{fri17}
B.~Friess, Y.~Peng, B.~Rosenow, F.~von Oppen, V.~Umansky, K.~von Klitzing, J.H.
  Smet, Nat Phys \textbf{13}, 11 (2017)

\bibitem{Lan18}
J.~Lane, L.~Zhang, M.~Khasawneh, B.~Zhou, E.~Henriksen, J.~Pollanen,
  arXiv:1801.05270  (2018)

\bibitem{Gus14}
M.~Gustafsson, T.~Aref, A.~Kockum, M.~Ekstr{\"o}m, G.~Johansson, P.~Delsing,
  Science \textbf{346}, 207 (2014)

\bibitem{Man17}
R.~Manenti, A.~Kockum, A.~Patterson, T.~Behrle, J.~Rahamim, G.~Tancredi,
  F.~Nori, P.~Leek, Nature Communications \textbf{8}, 975 (2017)

\bibitem{Arz67}
R.~Arzt, E.~Salzmann, K.~Dransfeld, Applied Physics Letters \textbf{10}(5), 165
  (1967)

\bibitem{Tok79}
M.~Tokumura, F.~Akao, Physics Letters \textbf{72A}(2), 131 (1979)

\bibitem{Che82}
J.~Cheeke, P.~Morisseau, Journal of Low Temperature Physics \textbf{46}(3/4),
  319 (1982)

\bibitem{Aok03}
Y.~Aoki, Y.~Wada, Y.~Sekimoto, W.~Yamaguchi, A.~Ogino, M.~Saithoh, R.~Nomura,
  Y.~Okuda, Journal of Low Temperature Physics \textbf{134}(3/4), 945 (2003)

\bibitem{Aok032}
Y.~Aoki, Y.~Sekimoto, Y.~Wada, W.~Yamaguchi, R.~Nomura, Y.~Okuda, Physica B
  \textbf{329-333}, 116 (2003)

\bibitem{Mor80}
P.~Morisseau, J.~Cheeke, Ultrasonics Symposium IEEE pp. 1040--1043 (1980)

\bibitem{Che822}
J.~Cheeke, P.~Morisseau, M.~Poirier, Physics Letters \textbf{88A}(7), 359
  (1982)

\bibitem{Kos90}
Y.~Kosevich, E.~Syrkin, Journal of Physics: Condensed Matter \textbf{2}, 5047
  (1990)

\bibitem{And97}
E.Y. Andrei (ed.), \emph{Two-Dimensional Electron Systems on Helium and other
  Cryogenic Substrates}, vol.~19 (Kluwer Academic Publishing, 1997)

\bibitem{Mon04}
Y.~Monarkha, K.~Kono, \emph{Two-Dimensional Coulomb Liquids and Solids}
  (Springer, Berlin, 2004)

\bibitem{Pob92}
F.~Pobell, \emph{Matter and Methods at Low Temperatures}, 3rd edn. (Springer,
  Berlin, 2007)

\bibitem{Dran70}
K.~Dransfeld, E.~Salzmann, \emph{Physical Acoustics}, vol.~7 (Academic Press,
  New York, 1970)

\bibitem{Pol09}
J.~Pollanen, H.~Choi, J.~Davis, B.~Rolfs, W.~Halperin, J. Phys. Con. Ser.
  \textbf{150}, 012037 (2009)

\bibitem{Har16}
G.~Harris, D.~McAuslan, E.~Sheridan, Y.~Sachkou, C.~Baker, W.~Bowen, Nature
  Physics \textbf{12}, 788 (2016)

\bibitem{Lor17}
L.D. Lorenzo, K.~Schwab, Journal of Low Temperature Physics \textbf{186}, 233
  (2016)

\bibitem{Sou17}
F.~Souris, H.~Christiani, J.~Davis, Applied Physics Letters \textbf{111},
  172601 (2017)

\bibitem{Chi17}
L.~Childress, M.~Schmidt, A.~Kashkanova, C.~Brown, G.~Harris, A.~Aiello,
  F.~Marquardt, J.~Harris, Phys. Rev. A \textbf{96}, 063842 (2017)

\end{thebibliography}

\end{document}